# A Non-Blind Watermarking Scheme for Gray Scale Images in Discrete Wavelet Transform Domain using Two Subbands

Abdur Shahid[1], Shahriar Badsha[2], Md. Rethwan Kabeer[3], Junaid Ahsan[4] and Mufti Mahmud[5]

[1] Department of Computer Science and Engineering, Chittagong University of Engineering and Technology
Chittagong-4349, Bangladesh

[2] Department of Electrical Engineering, University of Malaya
Kuala Lumpur, Malaysia

[3] Department of Computer Science and Information System, University of Malaya
Kuala Lumpur, Malaysia

[4] Department of Computer Science and Engineering, Chittagong University of Engineering and Technology
Chittagong-4349, Bangladesh

[5] Faculty of Business Administration, American International University-Bangladesh
Dhaka-1213, Bangladesh

**Abstract**
Digital watermarking is the process to hide digital pattern directly into a digital content. Digital watermarking techniques are used to address digital rights management, protect information and conceal secrets. An invisible non-blind watermarking approach for gray scale images is proposed in this paper. The host image is decomposed into 3-levels using Discrete Wavelet Transform. Based on the parent-child relationship between the wavelet coefficients the Set Partitioning in Hierarchical Trees (SPIHT) compression algorithm is performed on the LH3, LH2, HL3 and HL2 subbands to find out the significant coefficients. The most significant coefficients of LH2 and HL2 bands are selected to embed a binary watermark image. The selected significant coefficients are modulated using Noise Visibility Function, which is considered as the best strength to ensure better imperceptibility. The approach is tested against various image processing attacks such as addition of noise, filtering, cropping, JPEG compression, histogram equalization and contrast adjustment. The experimental results reveal the high effectiveness of the method.

## 1. Introduction

The phenomenal development of the World Wide Web in recent years has led to a strong demand for reliable and secure copyright protection techniques for digital data. Among various techniques, digital watermarking has been cited as an effective solution to protect multimedia data. Watermarking is a technique to embed information into a digital content imperceptibly. The embedded watermark gives the host digital content a unique, digital identity that can be used for a variety of applications. Watermarking for broadcast monitoring allows the advertisers of an advertisement to know if another station pirated the advertisement that has their watermark. Copy control watermarking technique doesn't allow copying of its content. Another application is Content authentication which is done by adding a watermark to digital works, photographs, surveillance camera videos as well as important scanned documents [1]. A watermarking system has three characteristics: First, the system is robust. The watermark in intentional and accidental image processing and transformation must not be deleted. Second, it is secure. The watermark should be accessible only by authorized users. Third, it is invisible. One may not be able to distinguish between the watermarked image and the original image.

A large number of watermarking systems address the problem of implementing invisible robust watermarks. Watermarking algorithms can be categorized based on the requirements for watermark extraction or detection. Non-blind watermarking schemes require both watermark and the original image for extraction. The semi-blind schemes require the watermark and secret key, whereas the blind schemes need only the secret key.

On the basis of embedding data in digital content, watermarking algorithms are divided into two groups of spatial and frequency domain. Most of the earlier watermarking systems were in spatial domain. These algorithms embed the watermark by directly modifying selected intensity or color values of the host image [2]. Embedding watermark in least significant bit (LSB) of selected intensity or color value is the simplest and popular spatial-domain technique. The main advantage of Spatial-domain methods is that they are more consistent with HVS model. But these methods are even vulnerable to simple image processing and lossy compression. On the other hand, frequency-domain watermarking techniques involve in transforming an image into frequency-domain coefficients and then some selected coefficients are modified to embed watermark. Discrete Fourier Transform (DFT), Discrete Cosine Transform (DCT), Discrete Wavelet Transform (DWT) etc are widely used in frequency-domain watermarking. Higher capacity and greater robustness against various attacks are the main features of frequency-domain watermarking.

In this paper, a non-blind watermarking scheme is proposed which operates in frequency domain. 3 level Discrete Wavelet Transform is performed on the host image and LH2 and HL2 subband coefficients are used to embed watermark. The proposed method is different than many other approaches because instead of using pseudorandom number sequence, a monochrome watermark image is used as watermark. The significant coefficients are selected by applying SPIHT algorithm on the DWT image. The maximum allowable distortion of selected coefficient is computed using Noise Visibility Function (NVF). The proposed approach has following advantages: 1) the SPIHT compression algorithm makes it highly resistive against large scale compression; cropping and many other image processing attacks whereas many watermarking approaches fall against such attacks and 2) the NVF gives it high imperceptibility than many other DWT based schemes.

The paper is organized as follows: section 2 focuses some related works. Section 3 gives some background discussions in brief. The proposed watermark embedding and extraction model is discussed in section 4. Section 5 contains the experimental results. The comparison of the proposed method with an existing method is shown in section 6 and section 7 encloses the conclusion.

## 2. Related Works

Several methods used DCT in frequency domain watermarking [2][3][4]. Cox [2] first proposed a DCT based spread spectrum watermarking algorithm that inserts a pseudorandom sequence as watermark into the perceptually highest significant DCT coefficients of the transformed image. The work is treated as a milestone in the field of digital watermarking. The application of spread spectrum technique to embed watermark in frequency domain made it highly robust because it spreads the watermark in all the pixels. As a result it is very tough to detect and remove the watermark [5][6]. Tsui [7] provided a method for embedding a watermark into color images using Fourier Transform.

On the other hand, due to its excellent time-frequency domain localization properties, the DWT has become an excellent method for solving difficult problems in image processing. DWT decomposes images into subbands of different scales and positions and consequently reconstructs them with high exactitude. Because of its greater robustness and higher imperceptibility, DWT has become very popular in watermarking [8][9][10][11][12][13][14]. Hsieh [6] proposed an approach by using the concept of Zerotree from Embedded Zerotree Wavelet (EZW) coding introduced by Shapiro [15]. In EZW, wavelet coefficient at a given scale is considered insignificant with respect to a given threshold. Most of the coefficients in these sub bands may have very small magnitudes and thus low energy, so few significant coefficients are exploited in EZW to give an efficient coding scheme. More improvement of EZW is achieved by Set Partitioning in Hierarchical Trees (SPIHT) algorithm proposed by Said [16]. In this method, more zerotrees are efficiently found and represented by separating the tree root from the tree. The SPIHT-based method has therefore become the core technology of the emerging multimedia standards MPEG-4 and JPEG 2000. Zhang [13] proposed a watermarking capacity analysis in DWT domain based on Noise Visibility Function (NVF). The projected DWT based watermarking by Yanbin [12] used NVF to modulate the selected coefficients.

## 3. Preliminaries

In this section, the concepts of Discrete Wavelet Transform (DWT), Set Partitioning in Hierarchical Trees (SPIHT) algorithm and Noise Visibility Function (NVF) are discussed briefly.

### 3.1 Discrete Wavelet Transform (DWT)

From its foundation, the Discrete Wavelet Transform (DWT) has become very popular in image processing. Its multi-resolution (MR) property infers an image into a hierarchical framework, where an image is decomposed into a set of resolutions. This property makes the transform superior than others because features that are not detected at one resolution may be easily detected at another. DWT uses a scaling function to create approximations of an image and a wavelet function to encode the information difference between adjacent approximations. The approximations are logarithmically spaced in frequency domain. Two-dimensional wavelet transform can be treated as a one-dimensional wavelet transform performed along the x and y axis. The 2d DWT requires a scaling function and three two-dimensional wavelets. The scaling function

$$\varphi(x,y) = \varphi(x)\varphi(y) \quad (1)$$

And the three wavelets are

$$\psi^H(x,y) = \psi(x)\psi(y) \quad (2)$$

$$\psi^V(x,y) = \psi(x)\psi(y) \quad (3)$$

$$\psi^D(x,y) = \psi(x)\psi(y) \quad (4)$$

These wavelets measures the intensity variations in different directions: $\psi^H$ measures variations in horizontal edges, $\psi^V$ measures variations along vertical edges and $\psi^D$ measures variations along diagonals. The scaled and translated basis functions

$$\varphi_{j,m,n}(x,y) = 2^{j/2}\varphi(2^j x - m, 2^j y - n) \quad (5)$$

$$\varphi^i_{j,m,n}(x,y) = 2^{j/2}\psi^i(2^j x - m, 2^j y - n)$$
$$i = \{H, V, D\} \quad (6)$$

Here H, V and D refer to the horizontal, vertical and diagonal wavelets respectively. The DWT of a function of size $M \times N$ is then

$$W_\varphi(j_0, m, n) = \frac{1}{\sqrt{MN}} \sum_{x=0}^{M-1}\sum_{y=0}^{N-1} f(x,y)\varphi_{j_0,m,n}(x,y) \quad (7)$$

$$W^i_\psi(j, m, n) = \frac{1}{\sqrt{MN}} \sum_{x=0}^{M-1}\sum_{y=0}^{N-1} f(x,y)\psi^i_{j,m,n}(x,y) \quad (8)$$

The $W_\varphi(j_0, m, n)$ is the approximation in starting scale $j_0$. $W^i_\psi(j, m, n)$ defines the horizontal, vertical and diagonal details at scales $j \geq j_0$. Where $j_0 = 0$, $j = 0,1,2,\ldots,j-1$ and $m = n = 0,1,2,\ldots,2^j - 1$. Using the Eq. (7) and (8), the is $f(x,y)$ obtained by the inverse DWT

$$f(x,y) = \frac{1}{\sqrt{MN}} \sum_m \sum_n W_\varphi(j_0, m, n)\varphi_{j_0,m,n}(x,y)$$

$$+ \frac{1}{\sqrt{MN}} \sum_{i=H,V,D} \sum_{j=j_0}^{\infty} \sum_m \sum_n W^i_\psi(j, m, n)\psi^i_{j,m,n}(x,y) \quad (9)$$

When a 1-level DWT is applied, the image is decomposed into four parts of high, middle and low frequencies-LL1, HL1, LH1 and HH1 subbands. The subbands labeled HL1, LH1 and HH1 represent the finer scale wavelet coefficients. The LL1 subband contains the approximate image.

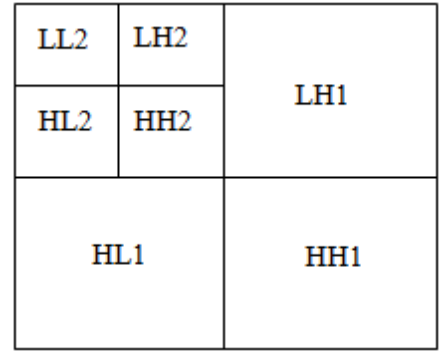

Fig. 1 Decomposition of DWT

A 3-level decomposition of 512×512 Lena image is shown in Fig 2.

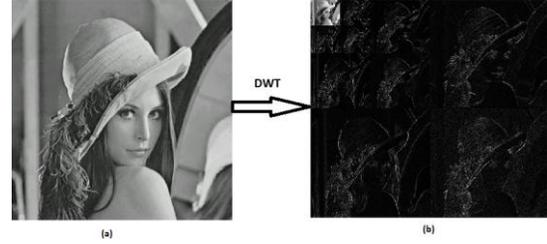

Fig. 2 DWT transformation of Lena image

3.2 SPIHT Algorithm

There is a parent-child relationship between the wavelet coefficients. Every coefficient at a given scale is related to a set of coefficients at the next finer scale of similar orientation. The coefficient at a coarse scale is called parent. Each parent has four children at the next finer scale of similar orientation.

Shapiro [15] introduced EZW compression algorithm which is based on the theory of parent-child relationship between the wavelet coefficients. It is simple, efficient and does not require any prior knowledge of original image. Its fine scalability and high effectiveness in removing spatial redundancy across the scales gave a breakthrough in image coding field. The SPIHT algorithm is an improvement to the EZW [16], which has achieved high performance with its combined properties of high speed, low complexity and easy applicability.

The encoding process of SPIHT consists of two quantization passes, the sorting pass and the refinement pass. The data structure of the algorithm consists of three linked lists, the LSP (list of significant pixels), the LIP (list of insignificant pixels), and the LIS (list of insignificant sets). These three lists are used to keep track of the elements of image during encoding. During sorting pass new significant entries in LIP and LIS are indentified and their signs are coded. In each refinement pass each coefficient in LSP except the ones added in the last sorting pass in refined. The image is reconstructed by the quantization process. The quantization step halves the threshold each time. The encoding process stopped when a target bit rate or threshold or quality requirement is reached.

The important definitions used in SPIHT algorithm are:
   (i,j): Wavelet coefficient at i th row and j th column.
   O(i,j): Set of offspring of the coefficient (i,j).
   O(i,j) = {(2i,2j),(2i,2j+1),(2i+1,2j),(2i+1,2j+1)}
In this algorithm, three linked lists are used to store the significance information during set partitioning. These are:
   LIP: List of insignificant coefficients.
   LSP: List of significant coefficients.
   LIS: List of insignificant sets.

Algorithm: SPIHT

Let,
The size of each of LH2 and HL2 = M×M
Band of each (i,j) coefficient is k.
1) Initialization:
   LSP = φ.
   LIP = φ
   LIS = (i,j) ϵ LH3 & HL3.
   Tlm = {(1/M) × mean (|LH2|), (1/M) × mean (|HL2|)}
   n = log2 max $_{(i,j)\ \epsilon\ LH3,\ HL3,\ LH2\ \&\ HL2}${|c(i,j)|}
   T = 2n
2) Sorting Pass:
   For each (i,j) ϵ LIP do
       If |c(i,j)|>=T & T>=Tlm(k) then
           Move (i,j) to LSP.
       End if
   End for
   For each (i,j) ϵ LIS do
       If |O(i,j)|>=T & T>=Tlm(k) then
           For each (k,l) ϵ O(i,j) do
               If |(k,l)|>=T & T>=Tlm(k)
                   Append (k,l) to LSP.
               Else
                   Append (k,l) to LIP.
               End if
           End for
       End if
   End for
3) n = n-1, T = 2^n.
4)    If T >=min(Tlm) ,goto step 2)

3.3 Noise Visibility Function (NVF)

Noise Visibility Function (NVF) is a human visual system (HVS) treated as the best method to establish connection between watermarking capacity and content of image. If we assume the original image to be subject to a generalized Gaussian distribution, the NVF of each wavelet coefficient is:

$$NVF(i,j) = \frac{1}{1+\sigma_x^2(i,j)} \quad (10)$$

$$\sigma_x^2 = \frac{1}{(2L+1)^2} \sum_{m=-L}^{L} \sum_{n=-L}^{L} (x(i+m, j+n) - \mu(i,j))^2 \quad (11)$$

$$\mu(i,j) = \frac{1}{(2L+1)^2} \sum_{m=-L}^{L} \sum_{n=-L}^{L} x(i+m, j+n) \quad (12)$$

Where, $x(i,j)$ is the value of $(i,j)$, $\sigma_x^2$ is the local variance of the $(i,j)$ centered local image area. Watson [17] presented a quantization matrix for the perceptually lossless compression in DWT domain. The quantization matrix is composed by the quantization factor of each wavelet levels and orientations.

Table 1: Quantization factors for four-level DWT [17]

| *Orientation* | *Level 1* | *Level 2* | *Level 3* | *Level 4* |
|---|---|---|---|---|
| 1 | 14.049 | 11.106 | 11.363 | 14.5 |
| 2 | 23.028 | 14.608 | 12.707 | 14.156 |
| 3 | 58.756 | 28.408 | 19.54 | 17.864 |
| 4 | 23.028 | 14.685 | 12.707 | 14.156 |

The maximum allowable distortion of a wavelet coefficient is:

$$\Delta(i,j) = (1 - NVF(i,j)).Q_{\lambda,\theta} + NVF(i,j).s_1 \quad (13)$$

Where, quantization factor $Q_{\lambda,\theta}$ is found from quantization matrix, $\lambda$ and $\theta$ are the wavelet level and orientation and s1 is the maximum allowable distortion in the flat region of the image.

3.4 Pretreatment of the watermark

The binary pattern watermark image is transformed into a one-dimensional vector. Then this vector is mapped to a vector for the effective embedding of watermark. The mapping formula is as follows,

$$W'(i) = \begin{cases} -1 & w(i) = 0 \\ 1 & w(i) = 1 \end{cases} \quad (14)$$

## 4. Proposed Watermarking System

This section discusses the proposed model to embed and extract the watermark for gray scale images.

4.1 Watermark Embedding Method

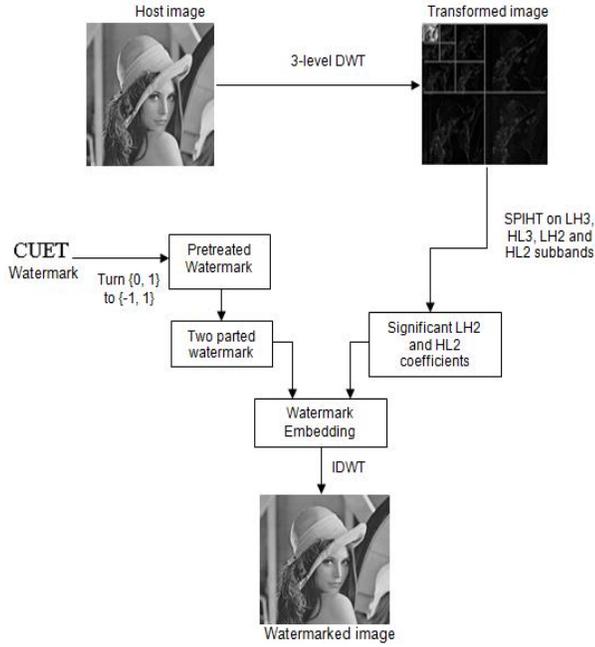

Fig. 3 Watermark embedding method

The embedding steps are:
1) The original watermark is transformed into a vector, then mapped from {0, 1} to {-1, 1} and yields the watermark vector, $W'$.
2) Divide $W'$ into two parts: $W' = W_1' + W_2'$ (15)
3) 3–level Discrete Wavelet Transform (DWT) is performed on the host image $I$ to decompose it into 10 subbands.
4) Apply SPIHT algorithm on LH3, LH2, HL3 and HL2 subbands.
5) Select the highest $(P \times Q)/2$ significant coefficients from each of LH2 and HL2 subbands. Here $P \times Q$ is the size of the watermark.
6) The maximum allowable distortions of the selected significant coefficients are computed.
7) $W'$ is embedded into the selected coefficients by using the following equation:

$$f'(x,y) = f(x,y) + \alpha \times w(i) \times \Delta(x,y) \quad (16)$$

Where $f(x,y)$ is the original wavelet coefficient, $f'(x,y)$ is the embedded coefficient, $\alpha$ is the scaling factor, $\Delta(x,y)$ is the maximum allowable distortion of $(x,y)$ and $W'(i)$ is the i-th bit of $W'$.
8) Inverse Discrete Wavelet Transform (IDWT) is performed and yields the watermarked image $I'$.

4.2 Watermark Extraction Method

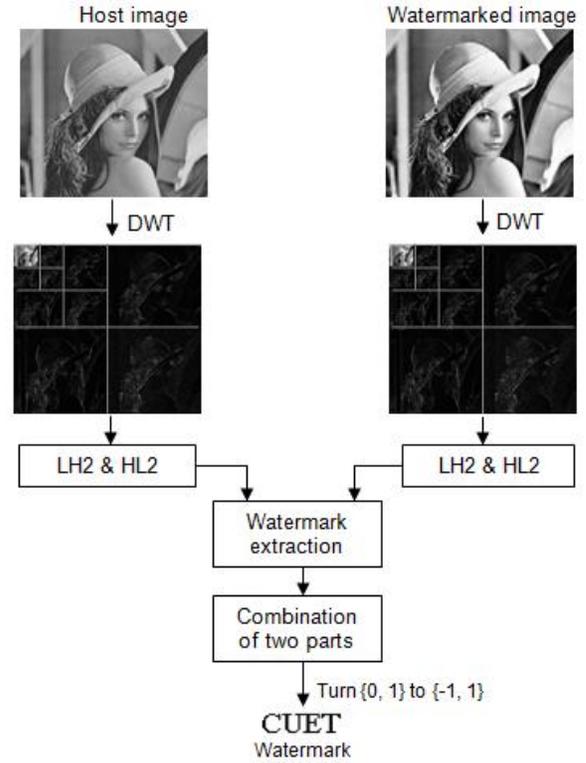

Fig. 4 Watermark extraction method

The watermark extraction steps are:
1) 3-level DWT is performed on both the original and watermarked (possibly attacked) image.
2) Embedding positions are retrieved from LH2 and HL2 subbands of the transformed images.
3) Extract watermark information from using the retrieved positions by using the following formula,

$$\overline{W}_k' = (I' - I)/(\alpha \times \Delta) \quad (17)$$

Here k is one of the subbands.
4) Combine the results of step 3 to obtain the extracted watermark:

$$\overline{W}' = \overline{W}_1' + \overline{W}_2' \quad (18)$$

5) Turn {-1, 1} in $\overline{W}^{'}$ to {0, 1} and get watermarking information $\overline{\overline{W}}$.

## 5. The Experimentation

The proposed approach is tested for more than 100 gray scale images. The size of each image is $512 \times 512$. A $32 \times 32$ binary image with "CUET" characters is used as watermark image. The following parameters are used to evaluate the performance of the proposed scheme.

1) **Correlation coefficient:** The correlation of the original watermark, $W$ and the extracted watermark, $\overline{\overline{W}}$ is calculated by using the following equation:

$$Correlation = \frac{\sum(x-\bar{x})(y-\bar{y})}{\sqrt{\sum(x-\bar{x})^2}\sqrt{\sum(y-\bar{y})^2}} \quad (19)$$

2) **Error metrics:** Mean Square Error (MSE) and Peak Signal to Noise Ratio (PSNR) are the two mostly used error metrics to compare various images. The MSE is the cumulative squared error between the original image and the modified image, whereas PSNR is the measure of the peak error. The formulas for the two are:

$$MSE = \frac{1}{MN}\sum_{y=1}^{M}\sum_{x=1}^{N}[I(x,y)-I^{'}(x,y)]^2 \quad (20)$$

$$PSNR = 20 \times \log_{10}(\frac{255}{\sqrt{MSE}}) \quad (21)$$

Where, $I(x,y)$ is the original image, $I^{'}(x,y)$ is the modified image and $M, N$ are the dimensions of the images. A lower value for MSE means lesser error, and as seen from the inverse relation between the MSE and PSNR, this translates to a high value of PSNR.

5.1 Test and watermark images

Fig. 5 and 6 show some test and watermark image used in this work respectively.

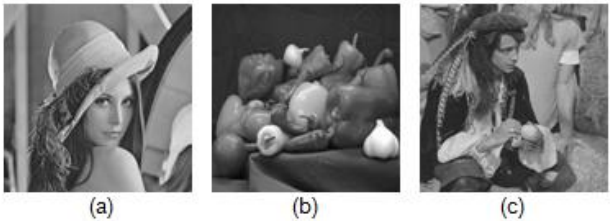

Fig. 5 Test images: (a) Lena, (b) Pepper and (c) Pirate

Fig. 6 The watermark image

Fig. 7 and 8 demonstrate the test images after embedding the watermark and extracted watermark from these watermarked images respectively.

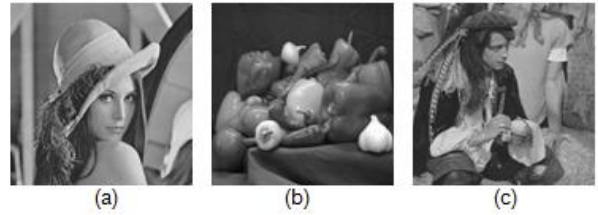

Fig. 7 The watermarked test images

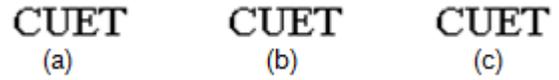

Fig. 8 The extracted watermarks from: (a) Lena, (b) Pepper, (c) Pirate

Table 2 shows the PSNR of the watermarked images and correlation coefficient between the original and extracted watermark.

Table 2: PSNR and correlation coefficient

| *Images* | *PSNR(dB)* | *Correlation Coefficient* |
|---|---|---|
| Lena | 48.0429 | 1 |
| Pepper | 48.0624 | 1 |
| Pirate | 49.6520 | 1 |

The scaling factor, for LH2 and HL2 subbands are:

Table 3: Scaling factor

| *Subband* | *Scaling factor* |
|---|---|
| LH2 | 3 |
| HL2 | 1 |

5.2 Results

The robustness of the watermarking algorithm is tested by applying various attacks on watermarked images such as salt & pepper noise and Gaussian noise addition, Gaussian, Mean and Median filtering, JPEG compression, Cropping, Contrast Adjustment and Histogram equalization.

1) **Addition of Noise:** In this work two types of noise are added to watermarked images: Salt & Pepper noise and Gaussian noise. Salt & Pepper noise is added to the watermarked test images with varying density. The watermarked images are also attacked by Gaussian noise with different noise variances. The correlation coefficients for the watermarked test images by applying Salt & Pepper noise and Gaussian noise are shown in table 4 and 5 respectively.

Table 4: Correlation coefficient for salt & pepper noise attack

| *Noise Density* | *Lena* | *Pepper* | *Pirate* |
|---|---|---|---|
| 0.01 | 0.9293 | 0.9130 | 0.9011 |
| 0.02 | 0.9121 | 0.9040 | 0.8960 |
| 0.03 | 0.8920 | 0.8890 | 0.8876 |

Table 5: Correlation coefficient for Gaussian noise attack

| *Variance* | *Lena* | *Pepper* | *Pirate* |
|---|---|---|---|
| 0.01 | 0.9439 | 0.9345 | 0.9566 |
| 0.02 | 0.9006 | 0.9063 | 0.8891 |
| 0.03 | 0.8775 | 0.8179 | 0.8118 |

2) **Filtering:** In this work Gaussian, Mean and Median filters with varying kernel size are applied on the test images to analyze the effectiveness of the scheme.

Table 6: Correlation coefficient for filtering attack

| *Filter* | *Lena* | *Pepper* | *Pirate* |
|---|---|---|---|
| Mean 3×3 | 0.8717 | 0.9456 | 0.7995 |
| Mean 5×5 | 0.6151 | 0.6227 | 0.6151 |
| Median 3×3 | 0.9839 | 0.9785 | 0.9456 |
| Median 5×5 | 0.8075 | 0.7861 | 0.6668 |
| Gaussian 3×3 | 1 | 1 | 1 |
| Gaussian 5×5 | 1 | 1 | 1 |

3) **Cropping:** Cropping is a pixel removal operation from a digital image. This lossy operation is treated as one of the most dreadful attacks for image watermarking. It is done by using deleting or zeroing rows or columns.

Table 7: Correlation coefficient for cropping attack

| *Cropped Area size* | *Lena* | *Pepper* | *Pirate* |
|---|---|---|---|
| 64×64 | 0.9400 | 1 | 1 |
| 128×128 | 0.9233 | 0.9946 | 0.9785 |

4) **JPEG compression:** Compression reduces the irrelevance and redundancy of the image for efficient transmission or reduces store size. The test images are tested with lossy JPEG compression with varying quality factor.

Table 8: Correlation coefficient for JPEG compression

| *Quality Factor (%)* | *Lena* | *Pepper* | *Pirate* |
|---|---|---|---|
| 70 | 1 | 1 | 1 |
| 50 | 1 | 1 | 1 |
| 30 | 1 | 1 | 0.9946 |
| 20 | 1 | 1 | 0.9785 |
| 10 | 0.9566 | 0.8361 | 0.9120 |

5) **Histogram Equalization:** Histogram equalization is one of the well known image enhancement techniques. It adjusts image intensities to enhance the contrast of that image.

Table 9: Correlation coefficient for histogram equalization

| *Lena* | *Pepper* | *Pirate* |
|---|---|---|
| 0.9511 | 0.9176 | 0.9511 |

6) **Contrast Adjustment:** The contrast adjustment process adjusts the histogram of an image by linearly scaling the pixel values between the upper and lower limits. Pixel values that are above or below this range are saturated to the upper or lower limit value, respectively.

Table 10: Correlation coefficient for contrast adjustment attack

| *Contrast Adjust (%)* | *Lena* | *Pepper* | *Pirate* |
|---|---|---|---|
| 10 | 0.9893 | 0.9893 | 0.9839 |
| 20 | 0.9773 | 0.9621 | 0.9511 |
| 30 | 0.9533 | 0.9421 | 0.9120 |

## 6. Comparison with existing method

Table 11: Robustness comparison

| *Attack* | *Proposed* | *Scheme in [14]* |
|---|---|---|
| Mean Filtering (13×13) | 0.1427 | -0.3696 |
| Median Filtering | 0.1227 | -0.3233 |
| (13×13) | | |
| Gaussian Noise (75%) | 0.4806 | 0.2843 |
| JPEG compression (Q.F:20) | 1 | 0.9922 |
| Histogram Equalization | 0.9176 | 0.8620 |
| Cropping((1/4)th area remaining) | 0.5095 | 0.3840 |
| Contrast Adjustment (50% increased) | 0.8306 | 0.7557 |

From the above comparisons, we can see that, the proposed method is better than the scheme in [14] in case of imperceptibility as well as robustness against various types of attack.

## 7. Conclusion

This paper has introduced a new robust and effective image watermarking method to embed a binary watermark image which takes into account the parent-child relationship between the wavelet coefficients. The proposed method is non-blind in nature, i.e. it requires the original test image at extraction phase. It is very much effective against JPEG compression, filtering, noise addition, contrast adjustment, histogram equalization and cropping attack. The significance of the proposed method relies on the combination of SPIHT compression algorithm and Noise Visibility Function (NVF). The SPIHT algorithm finds out the significant wavelet coefficients to embed watermark which ensures superior robustness. The exploitation of NVF gives the proposed scheme better imperceptibility than many of the DWT based schemes.

**First Author** received the B.Sc degree in Computer Science and Engineering from Chittagong University of Engineering and Technology (CUET), Chittagong, Bangladesh in 2011. He is currently serving as a software engineer at Samsung Bangladesh R&D Center Ltd, Bangladesh. His research interest includes Digital Image Processing, Digital Watermarking, Artificial Intelligence and Natural Language Processing.

**Second Author** received B.Sc degree in Computer Science and Engineering from Chittagong University of Engineering and Technology (CUET), Chittagong, Bangladesh in 2011. Currently he is a research assistant in Electrical Engineering department at University of Malaya, Kuala Lumpur, Malaysia. His research interest includes Biomedical Image Processing and Multimedia Security.

**Third Author** received his B.Sc degree in Computer Science and Engineering from Chittagong University of Engineering and Technology, Chittagong, Bangladesh in 2011. Currently he is a research assistant in department of Computer Science and Information System at University of Malaya, Kuala Lumpur,

Malaysia. His research interest includes image processing, data mining, computer architecture and signal processing.

**Fourth Author** received his B.Sc degree in Computer Science and Engineering from Chittagong University of Engineering and Technology, Chittagong, Bangladesh in 2011. Currently he is serving as a Software Engineer at Samsung Bangladesh R&D Center Ltd, Bangladesh. He is interested in Bioinformatics and image processing.

**Fifth Author** received his Bachelor degree from Faculty of Business Administration, American International University-bangladesh (AIUB), Bangladesh. He worked at Grammen Phone, leading mobile phone operator in Bangladesh. His research interest includes data mining, financial computing, image processing and cloud computing.